\documentclass[a4paper,11pt]{article}
\usepackage{amsmath, amssymb, graphicx, amsfonts, fullpage}

\newtheorem{obs}{Observation}

\def\RR{\mathbb R}

\def\cC{\mathcal C}

\def\cG{\mathcal G}

\def\cP{\mathcal P}

\def\cV{\mathcal V}
\def\cX{\mathcal X}

\def\b1{\mathbf 1}

\def\bar{\overline}

\newcommand{\nop}[1]{}

\def\cY{{\mathcal Y}}
\def\cZ{{\mathcal Z}}

\newcommand{\qed}{\hfill$\square$\bigskip}
\newcommand{\raf}[1]{(\ref{#1})}
\newcommand{\proof}{\noindent {\bf Proof}.~~}

\newcommand{\hide}[1]{}

\newtheorem{theorem}{Theorem}

\newtheorem{lemma}{Lemma}
\newtheorem{claim}{Claim}
\newtheorem{corollary}{Corollary}

\title{On Computing the Vertex Centroid of a Polyhedron\thanks{
During part of this work the second author was supported by Graduiertenkolleg fellowship for PhD studies provided by Deutsche Forschungsgemeinschaft.}}
\author{Khaled Elbassioni\thanks{Max Planck Institut f\"ur Informatik, D-66123, Saarbr\"ucken, Germany}\\ elbassio@mpi-sb.mpg.de \and Hans Raj Tiwary\thanks{FR Informatik, Universit\"at des Saarlandes, D-66123, Saarbr\"ucken, Germany}\\ hansraj@cs.uni-sb.de}
\date{}
\begin{document}

\maketitle

\begin{abstract}
Let $\mathcal{P}$ be an $\mathcal{H}$-polytope in $\mathbb{R}^d$ with vertex set $V$. The vertex centroid is defined as the average of the vertices in $V$. We prove that computing the vertex centroid of an $\mathcal{H}$-polytope is \#P-hard. Moreover, we show that even just checking whether the vertex centroid lies in a given halfspace is already \#P-hard for $\mathcal{H}$-polytopes. We also consider the problem of approximating the vertex centroid by finding a point within an $\epsilon$ distance from it and prove this problem to be \#P-easy by showing that given an oracle for counting the number of vertices of an $\mathcal{H}$-polytope, one can approximate the vertex centroid in polynomial time. We also show that any algorithm approximating the vertex centroid to \emph{any} ``sufficiently'' non-trivial (for example constant) distance, can be used to construct a fully polynomial approximation scheme for approximating the centroid and also an output-sensitive polynomial algorithm for the Vertex Enumeration problem. Finally, we show that for unbounded polyhedra the vertex centroid can not be approximated to a distance of $d^{\frac{1}{2}-\delta}$ for any fixed constant $\delta>0$.
\end{abstract}

\section{Introduction}
An intersection of a finite number of closed halfspaces in $\RR^d$ defines a polyhedra. A polyhedra can also be represented as $conv(V)+cone(Y)$, the Minkowski sum of the convex hull of a finite set of points $V$ and the cone of a finite set of rays. A bounded polyhedra is called a polytope. In what follows, we will discuss mostly polytopes for simplicity and refer to the unbounded case explicitly only towards the end.

Let $\mathcal{P}$ be an $\mathcal{H}$-polytope in $\mathbb{R}^d$ with vertex set $V$. Various notions try to capture the essence of a ``center'' of a polytope. Perhaps the most popular notion is that of the center of gravity of $\mathcal{P}$. Recently Rademacher proved that computing the center of gravity of a polytope is \#P-hard \cite{DBLP:conf/compgeom/Rademacher07}. The proof essentially relies on the fact that the center of gravity captures the volume of a polytope perfectly and that computing the volume of a polytope is \#P-hard \cite{DBLP:journals/siamcomp/DyerF88}. Note that, polynomial algorithms exist that approximate the volume of a polytope within any arbitrary factor \cite{random_walk_kannan}. It is also easy to see that the center of gravity can be approximated by simply sampling random points from the polytope, the number of samples depending polynomially on the desired approximation (See Algorithm 5.8 of \cite{random_walk_kannan}).

In this paper we study a variant of the notion of ``center'' defined as the centroid (average) of the vertices of $P$. Despite being quite a natural feature of polytopes, this variant seems to have received very little attention both from theoretical and computational perspectives. Throughout this paper we will refer to the vertex centroid just as centroid. The reader should note that in popular literature the word centroid refers more commonly to the center of gravity. We nevertheless use the same terminology for simplicity of language. Our motivation for studying the centroid stems from the fact that the centroid encodes the number of vertices of a polytope. As we will see, this also makes computing the centroid hard.

The parallels between centroid and the center of gravity of a polytope mimic the parallels between the volume and the number of vertices of a polytope. Computing the volume and the number of vertices are both \#P-complete (\cite{counting_vertex_dyer, DBLP:journals/siamcomp/DyerF88, counting_vertices_linial}) and so are the problems of computing the corresponding centroids (\cite{DBLP:conf/compgeom/Rademacher07}, Theorem \ref{t1}). The volume can be approximated quite well but approximating the number of vertices of a polytope is an interesting open problem. Similarly, the center of gravity can be approximated quite well but (as we will see in this paper) obtaining a polynomial algorithm for approximating the centroid would be a very interesting achievement.

The problem of enumerating vertices of an $\mathcal{H}$-polytope has been studied for a long time. However, in spite of years of research it is neither known to be hard nor is there an output sensitive polynomial algorithm for it. A problem that is polynomially equivalent to the Vertex Enumeration problem is to decide if a given list of vertices of an $\mathcal{H}$-polytope is complete \cite{DBLP:journals/comgeo/WelzlABS97}. In this paper we show that any algorithm that approximates the centroid of an arbitrary polytope to any ``sufficiently'' non-trivial distance can be used to obtain an output sensitive polynomial algorithm for the Vertex Enumeration problem.

The main results of this paper are the following:
\begin{itemize}
 \item Computing the centroid of an $\mathcal{H}$-polytope is \#P-hard.
 \item Even just deciding whether the centroid of an $\mathcal{H}$-polytope lies in a halfspace remains \#P-hard.
 \item Approximating the centroid of an $\mathcal{H}$-polytope is \#P-easy.
 \item Any algorithm approximating the centroid of an arbitrary polytope within a distance $d^{\frac{1}{2}-\delta}$ can be used to obtain a fully polynomial approximation scheme for the centroid approximation problem and also an output sensitive polynomial algorithm for the Vertex Enumeration problem.
 \item There is no polynomial algorithm that approximates the vertex centroid of arbitrary $\mathcal{H}$-polyhedron within a distance $d^{\frac{1}{2}-\delta}$ for any fixed constant $\delta>0$, unless $P=NP$.
\end{itemize}

We should remark that for the approximation of centroid, we only consider polytopes (and polyhedra) whose vertices lie inside a unit hypercube. To see how this assumption can easily be satisfied, notice that a halfspace $h$ can be added to a polyhedron $P$ such that  $P\cap h$ is bounded and the vertices of $P$ are preserved in $P\cap h$. Also, such a halfspace can be found in polynomial time from the inequalities defining $P$. Once we have a polytope in $\RR^d$, solving $2d$ linear programs gives us the width along each coordinate axis. The polytope can be scaled by a factor depending on the width along each axis to obtain a polytope all whose vertices lie inside a unit hypercube. In case we started with a polyhedron $P$, the scaled counterpart of the halfspace $h$ that was added can be thrown to get back a polyhedron that is a scaled version of $P$ and all whose vertices lie inside the unit hypercube. In subsection \ref{ss2} we provide further motivation for this assumption.

Since all the vertices of the polytope (or polyhedron) lie inside a unit hypercube, picking any arbitrary point from inside this hypercube yield a $d^{\frac{1}{2}}$-approximation of the vertex centroid. Thus, the last result should be contrasted to the fact that approximating the vertex centroid within a distance of $d^\frac{1}{2}$ is trivial. Also, even though we discuss only polytopes \emph{i.e.} bounded polyhedra in subsections \ref{ss1} and \ref{ss2}, the results and the proofs are valid for the unbounded case as well. We discuss the unbounded case explicitly only in subsection \ref{ss3}.

%

\section{Results}\label{results}
\subsection{Exact Computation of the Centroid}\label{ss1}
The most natural computational question regarding the centroid of a polytope is whether we can compute the centroid efficiently. The problem is trivial if the input polytope is presented by its vertices. So we will assume that the polytope is presented by its facets. Perhaps not surprisingly, computing the centroid of an $\mathcal{H}$-polytope turns out be \#P-hard. We prove this by showing that computing the centroid of an $\mathcal{H}$-polytope amounts to counting the vertices of the same polytope, a problem known to be \#P-hard.

\begin{theorem}\label{t1}
 Given an $\mathcal{H}$-polytope $\mathcal{P}\subset \mathbb{R}^d$, it is \#P-hard to compute its centroid $c(\mathcal{P})$.
\end{theorem}
\proof
 Embed $\mathcal{P}$ in $\mathbb{R}^{d+1}$ by putting a copy of $\mathcal{P}$ in the hyperplane $x_{d+1}=1$ and making a pyramid with the base $\mathcal{P}$ and apex at the origin. Call this new polytope $\mathcal{Q}$. Treating the direction of the positive $x_{d+1}$-axis as up, it is easy to see that the centroid of the new polytope lies at a height $1-\frac{1}{n+1}$ iff the number of vertices of $\mathcal{P}$ is $n$. Thus any algorithm for computing the centroid can be run on $\mathcal{Q}$ and the number of vertices of $\mathcal{P}$ can be read off the $(d+1)$-st coordinate.
\qed

Suppose, instead, that one does not want to compute the centroid exactly but is just interested in knowing whether the centroid lies to the left or to the right of a given arbitrary hyperplane. This problem turns out to be hard too, and it is not difficult to see why.

\begin{theorem}\label{t2}
 Given an $\mathcal{H}$-polytope $\mathcal{P}\subset \mathbb{R}^d$ and a hyperplane $h=\{a\cdot x = b\}$, it is \#P-hard to decide whether $a\cdot c(\mathcal{P})\leq b$.
\end{theorem}
\proof
 Consider the embedding and the direction pointing upwards as used in the proof of Theorem \ref{t1}. Given an oracle answering sidedness queries for the centroid and any arbitrary hyperplane, one can perform a binary search on the height of the centroid and locate the exact height. The number of queries needed is only logarithmic in the number of vertices of $\mathcal{P}$.
\qed

\subsection{Approximation of the Centroid}\label{ss2}
As stated before, even though computing the gravitational centroid of a polytope exactly is \#P-hard, it can be approximated to any precision by random sampling. Now we consider the problem of similarly approximating the vertex centroid of an $\mathcal{H}$-polytope. Let $dist(x,y)$ denote the Euclidean distance between two points $x,y\in \mathbb{R}^d$. We are interested in the following problem:

\noindent
\textbf{Input:} $\mathcal{H}$-polytope $P\subset \mathbb{R}^d$ and a real number $\epsilon>0.$

\noindent
\textbf{Output:} $p\in\mathbb{R}^d$ such that $dist(c(P),p)\leq \epsilon.$

\medskip

We would like an algorithm for this problem that runs in time polynomial in the number of facets of $P$, the dimension $d$ and $\frac{1}{\epsilon}$. Clearly, such an algorithm would be very useful because if such an algorithm is found then it can be used to test whether a polytope described by $m$ facets has more than $n$ vertices, in time polynomial in $m,n$ and the dimension $d$ of the polytope by setting $\epsilon< \frac{1}{2}\left ( \frac{1}{n}-\frac{1}{n+1}\right ).$ This in turn would yield an algorithm that computes the number of vertices $n$ of a $d$-dimensional polytope with $m$ facets, in time polynomial in $m,n$ and $d$. As stated before, a problem that is polynomially equivalent to the Vertex Enumeration problem is to decide if a given list of vertices of an $\mathcal{H}$-polytope is complete \cite{DBLP:journals/comgeo/WelzlABS97}. Clearly then, a polynomial approximation scheme for the centroid problem would yield an output-sensitive polynomial algorithm for the Vertex Enumeration problem. 

Also, the problem of approximating the centroid is not so interesting if we allow polytopes that contain an arbitrarily large ball, since this would allow one to use an algorithm for approximating the centroid with \emph{any} guarantee to obtain another algorithm with an arbitrary guarantee by simply scaling the input polytope appropriately, running the given algorithm and scaling back. So we will assume that the polytope is contained in a unit hypercube in $\mathbb{R}^d$.

Now we prove that the problem of approximating the centroid is \#P-easy. We do this by showing that given an algorithm that computes the number of vertices of an arbitrary polytope (a \#P-complete problem), one can compute the centroid to any desired precision by making a polynomial (in $\frac{1}{\epsilon}$, the number of facets and the dimension of the polytope) number of calls to this oracle. Notice that in the approximation problem at hand, we are required to find a  point within a $d$-ball centered at the centroid of the polytope and radius $\epsilon.$ We first modify the problem a bit by requiring to report a point that lies inside a hypercube, of side length $2\epsilon$, centered at the centroid of the polytope. (The hypercube has a clearly defined center of symmetry, namely its own vertex centroid.) To see why this does not essentially change the problem, note that the unit hypercube fits completely inside a $d$-ball with the same center and radius $\frac{\sqrt{d}}{2}$. We will call any point that is a valid output to this approximation problem, an $\epsilon$-approximation of the centroid $c(P).$

Given an $\mathcal{H}$-polytope $P$ and a hyperplane $\{a\cdot x=b\}$ that intersects $P$ in the relative interior and does not contain any vertex of $P$, define $P_1$ and $P_2$ as follows:
\begin{eqnarray*}
 P_1&=&P\cap \{x|a\cdot x\leq b\}\\
 P_2&=&P\cap \{x|a\cdot x\geq b\}
\end{eqnarray*}

Let $V_1$ be the common vertices of $P_1$ and $P$, and $V_2$ be common vertices of $P_2$ and $P$. The following lemma gives a way to obtain the $\epsilon$-approximation of the centroid of $P$ from the $\epsilon$-approximations of the centroids of $V_1$ and $V_2$.

\begin{lemma}
 Given $P,V_1,V_2$ defined as above, let $n_1$ and $n_2$ be the number of vertices in $V_1$ and $V_2$ respectively. If $c_1$ and $c_2$ are $\epsilon$-approximations of the centroids of $V_1$ and $V_2$ respectively, then $c=\frac{n_1 c_1 + n_2 c_2}{n_1+n_2}$ is an $\epsilon$-approximation of the centroid $c^{*}$ of $P.$
\end{lemma}
\proof
 Let $c_{ij}$ be the $j$-th coordinate of $c_i$ for $i\in \{1,2\}$. Also, let $c^{*}_i$ be the actual centroid of $V_i$ with $c^{*}_{ij}$ denoting the $j$-th coordinate of $c^{*}_i$. Since $c_i$ approximates $c^{*}_i$ within a hypercube of side-length $2\epsilon$, for each $j\in \{1,\cdots,d\}$ we have 

\begin{eqnarray*}
 c^{*}_{ij} - \epsilon \leq &c_{ij}& \leq c^{*}_{ij} + \epsilon
\end{eqnarray*}
Also, since $c^{*}$ is the centroid of $P$, $$c^{*}=\frac{n_1 c^{*}_1+n_2 c^{*}_2}{n_1+n_2}$$

Hence, for each coordinate $c^{*}_j$ of $c^{*}$ we have
\begin{eqnarray*}
 & \frac{n_1 (c_{1j}-\epsilon)+n_2 (c_{2j}-\epsilon)}{n_1+n_2} & \leq  c^{*}_j  \leq  \frac{n_1 (c_{1j}+\epsilon)+n_2 (c_{2j}+\epsilon)}{n_1+n_2}\\
\Rightarrow & \frac{n_1 c_{1j}+n_2 c_{2j}}{n_1+n_2}-\epsilon  & \leq  c^{*}_j  \leq \frac{n_1 c_{1j}+n_2 c_{2j}}{n_1+n_2}+\epsilon\\
\Rightarrow & c_j - \epsilon & \leq  c^{*}_j  \leq c_j + \epsilon \\
\Rightarrow & c^{*}_j - \epsilon & \leq  c_j  \leq c^{*}_j + \epsilon
\end{eqnarray*}
\qed

Now to obtain an approximation of the centroid, we first slice the input polytope $P$ from left to right into $\frac{1}{\epsilon}$ slices each of thickness at most $\epsilon$. Using standard perturbation techniques we can ensure that any vertex of the input polytope does not lie on the left or right boundary of any slice. For each slice any point in the interior gives us an $\epsilon$-approximation of the vertices of $P$ that are contained in that slice. We can compute the number of vertices of $P$ lying in this slice by using the oracle for vertex computation and then using the previous Lemma we can obtain the centroid of $P$. Thus we have the following theorem:

\begin{theorem}
 Given a polytope $P$ contained in the unit hypercube, the $\epsilon$-approximation of the centroid of $P$ can be computed by making a polynomial number of calls to an oracle for computing the number of vertices of a polytope.
\end{theorem}

Now we present a bootstrapping theorem indicating that any ``sufficiently'' non-trivial approximation of the centroid can be used to obtain arbitrary approximations. For the notion of approximation let us revert back to the Euclidean distance function. Thus, any point $x$ approximating the centroid $c$ within a parameter $\epsilon$ satisfies $dist(x,c) \leq \epsilon$. As before we assume that the polytope $\mathcal{P}$ is contained in the unit hypercube. Since the polytope is thus contained in a hyperball with origin as its center and radius at most $\frac{\sqrt{d}}{2}$, any point inside $\mathcal{P}$ approximates the centroid within a factor $\sqrt{d}$. Before we make precise our notion of ``sufficiently'' non-trivial and present the bootstrapping theorem, some preliminaries are in order.

\begin{lemma}\label{l5}
 Suppose $(x,y),(u,u) \in \mathbb{R}^{2d}$, where $x,y,u \in \mathbb{R}^d$, then $$ ||u-\frac{x+y}{2}||\leq \frac{||(u,u)-(x,y)||}{\sqrt{2}},$$ where $||\cdot||$ is the Euclidean norm. 
\end{lemma}

The proof of the above lemma is easy and elementary, and hence we omit it here. Next, consider the product of two polytopes. Given $d$-dimensional polytopes $\mathcal{P},\mathcal{Q}$ the product $\mathcal{P}\times \mathcal{Q}$ is defined as the set $\{(x,y)|x\in \mathcal{P}, y \in \mathcal{Q}\}$. The facet defining inequalities of the product of $P,Q$ can be computed easily from the inequalities defining $P$ and $Q$. 
\begin{eqnarray*}
 P&=&\{x|A_1x\leq b_1 \},\\
 Q&=&\{y|A_2y\leq b_2 \}\\
 \Rightarrow P\times Q&=&\{(x,y)|A_1x\leq b_1, A_2y\leq b_2 \},\\
\end{eqnarray*}
where $A_1\in\RR^{m_1\times d_1}, A_2\in\RR^{m_2\times d_2},x\in \RR^{d_1},y\in \RR^{d_2}, b_1\in \RR^{m_1\times 1}, b_2\in \RR^{m_2\times 1}$.

It is easy to see that the number of vertices of $\mathcal{P}\times \mathcal{Q}$ is the product of the number of vertices of $\mathcal{P}$ and that of $\mathcal{Q}$, and the number of facets of $\mathcal{P}\times \mathcal{Q}$ is the sum of the number of facets of $\mathcal{P}$ and that of $\mathcal{Q}$. Moreover, the dimension of $\mathcal{P}\times \mathcal{Q}$ is the sum of the dimensions of $\mathcal{P}$ and that of $\mathcal{Q}.$

\begin{obs}
 If $c$ is the centroid of a polytope $P$ then $(c,c)$ is the centroid of $P\times P$.
\end{obs}

Suppose we are given an algorithm for finding $\epsilon$-approximation of an arbitrary polytope contained in the unit hypercube. For example, for the simple algorithm that returns an arbitrary point inside the polytope, the approximation guarantee is $\frac{\sqrt{d}}{2}$. We consider similar algorithms whose approximation guarantee is a function of the ambient dimension of the polytope. Now suppose that for the given algorithm the approximation guarantee is $f(d)$. For some parameter $k$ consider the $k$-fold product of $P$ with itself $\overbrace{P\times \cdots \times P}^{k\ times}$, denoted by $P^k$. Using the given algorithm one can find the $f(2^kd)$ approximation of $P^k$ and using Lemma \ref{l5} one can then find the $\frac{f(2^kd)}{\sqrt{2}^k}$-approximation of $P$. This gives us the following bootstrapping theorem:
\begin{theorem}\label{t4}
 Suppose we are given an algorithm that computes a $\frac{\sqrt{d}}{g(d)}$-approximation for any polytope contained in the unit hypercube in polynomial time, where $g(.)$ is an unbounded monotonically increasing function. Then, one can compute an $\epsilon$-approximation in time polynomial in the size of the polytope and $g^{-1}(\frac{\sqrt{d}}{\epsilon})$. 
\end{theorem}

In particular, if we have an algorithm with $d^{\frac{1}{2}-\delta}$ approximation  guarantee for finding the centroid of any polytope for some fixed constant $\delta>0$, then this algorithm can be used to construct a fully polynomial approximation scheme for the general problem.

\subsection{Approximating centroid of a polyhedron is hard}\label{ss3}

The reader should note that the analysis of subsections \ref{ss1} and \ref{ss2} remains valid even for the unbounded case (polyhedra). Even though we do not have any idea about the complexity of approximating the centroid of a polytope, now we show that for an arbitrary unbounded polyhedron the vertex centroid can not be $d^{\frac{1}{2}-\delta}$-approximated for any fixed constant $\delta>0$. To show this we first prove that for an $\mathcal{H}$-polyhedron $P\subset \RR^d$ the vertex centroid of $P$ can not be $\frac{1}{d}$-approximated in polynomial time unless $P=NP$. This together with Theorem \ref{t4} completes the proof for hardness of $d^{\frac{1}{2}-\delta}$-approximation of the centroid of an $\mathcal{H}$-polyhedron.

The proof sketch is as follows: Given a boolean CNF formula $\phi$, we construct a graph $G(\phi)$ such that $G(\phi)$ has a ``long'' negative cycle if and only if $\phi$ is satisfiable. For a given graph $G$ we define a polyhedron $P(G)$ such that every negative cycle in $G$ is a vertex of $P(G)$ and vice-versa. From the properties of the vertex centroid of this class of polyhedron, we then prove that for any formula $\phi$, $\frac{1}{d}$-approximating the vertex centroid of $P(G(\phi))$ would reveal whether $\phi$ is satisfiable or not.

Before we proceed further, we would like to remark that for the graph corresponding to a given 3-CNF formula, we use the same construction that appears in \cite{conf/soda/KhachiyanBBEG06} except for different edge weights. We nevertheless include the discussion here for completeness. Our main contribution is the negative-flow polyhedron associated with a directed graph and a set of edge weights and the analysis of the vertices of this polyhedron leading to Theorem \ref{t5}.

\subsubsection{Graph of a CNF formula}
Given a 3-CNF $\phi(x_1,\ldots,x_N) = C_1\wedge \ldots \wedge C_m,$ we construct a weighted directed graph $G=(V,E)$, on $|V|=5\sum_{j=1}^m|C_j|+m-n+1=16m-n+1$ vertices and $|E|=6\sum_{j=1}^m|C_j|+1=18m+1$ arcs (where $|C_j|$ denotes the number of literals appearing in clause $C_j$, \emph{i.e.} $3$ for a $3$-CNF), as follows. For each literal $\ell=\ell^j$ appearing in clause $C_j$, we introduce two paths of three arcs each: $\cP(\ell)=(p(\ell),a(\ell),b(\ell),q(\ell))$, and $\cP'(\ell)=(r(\ell),b'(\ell),a'(\ell),s(\ell))$. The weights of these arcs are set as follows:  $$
\begin{array}{lll}
w((p(\ell),a(\ell)))=\frac{1}{2}, & w((a(\ell),b(\ell)))=-\frac{1}{2}, & w((b(\ell),q(\ell)))=0,\\\\*[2mm]
w((r(\ell),b'(\ell)))=0, &w((b'(\ell),a'(\ell)))=-\frac{1}{2},& w((b'(\ell),s(\ell)))=\frac{1}{2}.
\end{array}
$$
           
\begin{figure}[btp] 
\center\includegraphics[width=15cm]{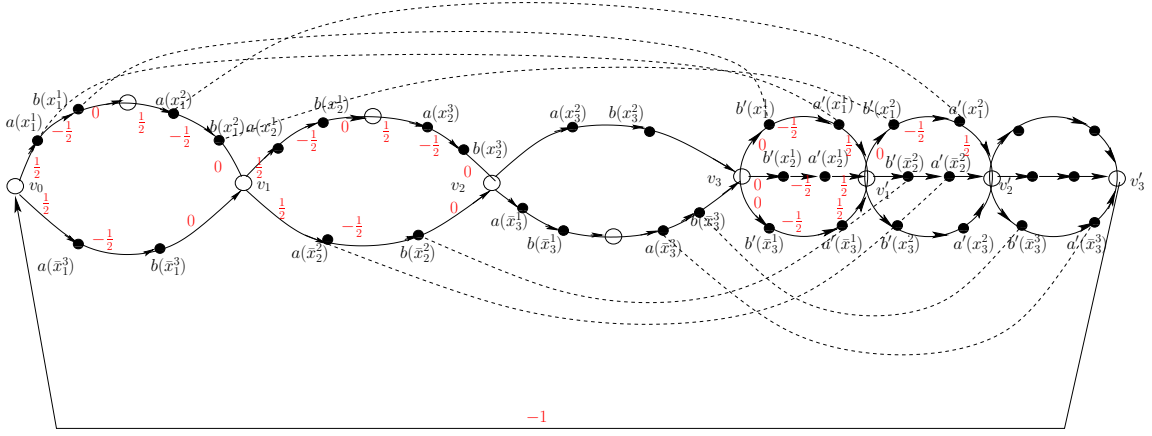}\\
\caption{An example of the graph construction for CNF $C=(x_1\vee
x_2\vee\overline{x}_3)\wedge(x_1\vee\overline{x}_2\vee x_3)\wedge(\overline{x}_1\vee x_2\vee\overline{x}_3)$. }
\label{f2}
\end{figure}

These edges are connected in $G$ as follows (see Figure \ref{f2} for an example):
$$G= v_0~\cG_1~v_1 ~ \cG_2 ~v_2 \ldots ~v_{n-1}~ \cG_n~v_n~\cG_1'~v_1'~\cG_2'~v_2'\ldots~v_{m-1}'~\cG_m'~v_m',$$
where $v_0,v_1,\ldots,v_{n},v_1',\ldots,v_{m-1}',v_m'  $ are
distinct vertices, each $\cG_i=\cY_i\vee\cZ_i$, for $i=1,\ldots,n$,
consists of two parallel chains $\cY_i=\wedge_{j}\cP(x_i^j)$
and $\cZ_i=\wedge_{j}\cP(\bar x_i^j)$ between $v_{i-1}$ and $v_i$,
and each $\cG_j'=\vee_{i=1}^{|C_j|}\cP'(\ell_i^j)$, for $j=1,\ldots,m$, where $\ell^j_1,\ell^j_2,\ldots$ are the literals appearing in $C_j$.

Finally we add the arc $(v_m',v_0)$ with weight $-1$, and \emph{identify} the pairs of nodes $\{a(\ell),a'(\ell)\}$ and $\{b(\ell),b'(\ell)\}$ for all $\ell$, (i.e. $a(\ell)=a(\ell')$ and $b(\ell)=b(\ell')$ define the same nodes).

We will argue now that all negative cycles of $G$ have weight $-1$. Clearly the arcs $(a(\ell),b(\ell))$ and $(b'(\ell),a'(\ell))$ form a directed cycle of total weight $-1$, for every literal occurrence $\ell$. There are $\sum_{j=1}^m|C_j|=3m$ such cycles, corresponding to a subset $\cX\subseteq\cV(P(G,w))$. 

Call a cycle of $G$ long if it contains the vertices $v_0,v_1,\ldots,v_{n},v_1',\ldots,v_{m-1}',v_m'$. Any long cycle has weight $-1$. The crucial observation is the following.
 
\begin{claim}\label{cl6}
Any negative cycle $C\in\cC^-(G,w)\setminus\cX$ must be long.
\end{claim}
\proof
Consider any cycle $C\not\in\cX$, and let us write the traces of the nodes visited on the cycle (dropping the literals, and considering $a,a'$ and $b,b'$ as different copies), without loss of generality as follows:
$$
p~a~b~p~a~b~p~\cdots~a~a'~s~b'~a'~s~b'~\cdots~b'~b~p~a~b~\cdots~p. 
$$
Note that the sequences $a'~a$ and $b~b'$ are not allowed since otherwise $C$ contains a cycle from $\cX$. 

Let us compute the distance (i.e., the total weight) of each node on this sequence starting from the initial $p$. Call the subsequences $a'~a'$ and $b~b'$, $a$- and $b$-jumps respectively. Then it is easy to verify that each $a$-jump causes the distance to eventually increase by $1$ while each $b$-jump keeps the distance at its value. More precisely, the distance at a node $x$ in the sequence is given by $d(x)=t+d_0-\delta(x)$, where $t$ is the number of $a$-jumps appearing upto $x$, and
$$
d_0=\left\{
\begin{array}{ll}
0&\mbox{if $x\in\{p,s\}$},\\
\frac{1}{2}&\mbox{if $x=a$},\\
-\frac{1}{2}&\mbox{if $x=a'$},\\
0&\mbox{if $x=b=b'$},\\
\end{array}
\right.
~~\delta(x)=\left\{
\begin{array}{ll}
1& \mbox{if arc $(v_m',v_0)$ appears on the path from $p$ to $x$}\\
0& \mbox{otherwise}. 
\end{array}
\right.
$$
One also observes that, if the sequence has a $b$-jump, then it must also contain an $a$-jump. Thus it follows from the definition of $d(x)$ that any cycle with a jump must be non-negative. So the only possible negative cycle not in $\cX$ must be long. 
\qed

If we are interested only in the negative cycles then it is clear that long cycles are exactly those negative cycles that contain the edge $(v_m',v_0)$. In what follows this edge will be called the \textbf{marked} edge, the negative cycles containing this marked edge will be called \textbf{long} negative cycles and the other negative cycles will be called \textbf{short} negative cycles.

By Claim \ref{cl6}, checking of $\cV(P(G,w))=\cX$ is equivalent to checking if $G$ has a long cycle. It is easy to see that the latter condition is equivalent to the non-satisfiability of the input CNF formula $\phi$ (see \cite{conf/soda/KhachiyanBBEG06} for more details). 

To summarize this subsection, for a $3$-CNF with $m$ clauses we can obtain a graph $G$ with $18m+1$ edges such that $G$ has $3m$ short negative cycles and finding whether $G$ has any negative cycle (called long negative cycle) containing one special edge is NP-complete. Furthermore, each of these negative cycles has total weight $-1$.

\subsubsection{The polyhedron of negative-weight flows of a graph}
Given a directed graph  $G = (V,E)$
and a weight function  $w: E \rightarrow \RR$ on its arcs, consider the following polyhedron:  
\[
P(G,w)=\left\{y\in \RR^E~\left|~
\begin{array}{cl}\displaystyle (F)~~~\sum_{v:(u,v)\in
E}y_{uv}-\sum_{v:(v,u)\in E}y_{vu} ~=~ 0 &\forall~~ u\in V\\*[7mm]
\displaystyle (N)~~~~~~~~~~~~~~~~\sum_{(u,v)\in E} w_{uv}y_{uv} ~=~ 
-1&\\*[3mm]y_{uv}\geq 0 &\forall~~
(u,v)\in E\end{array}\right.\right\}.
\]

If we think of $w_{u,v}$ as the cost/profit paid for edge $(u,v)$ per unit of flow, then each point of $P(G,w)$ represents a \emph{negative-weight circulation} in $G$, i.e., assigns a non-negative flow on the arcs, obeying the \emph{conservation of flow} at each node of $G$, and such that total weight of the flow is strictly negative. 

For a subset $X\subseteq E$, and a weight function $w:E\mapsto\RR$, we denote by $w(X)=\sum_{e\in X}w_e$, the total weight of $X$. For $X\subseteq E$, we denote by $\chi(X)\in\{0,1\}^E$ the characteristic vector of $X$: $\chi_e(X)=1$ if and only if $e\in X$, for $e\in E$. The following theorem states that the vertex set $\cV(P(G,w))$ of $P(G,w)$ is in one-to-one correspondence with the negative cycles of the graph $G$.

\begin{theorem}\label{t5}
Let $G=(V,E)$ be a directed graph and $w: E \rightarrow \RR$ be a 
real weight on the arcs. Then
\begin{eqnarray}
\label{e1}
\cV(P(G,w))&=&\left\{\frac{-1}{w(C)}\chi(C):~C\in\cC^-(G,w)\right\}.
\end{eqnarray}
\end{theorem}
We prove this theorem by proving a series of claims about the graph $G$. Let $m=|E|$ and $n=|V|$. It is easy to verify that any element $y\in\RR^E$ of the set on the right-hand side of \raf{e1} belongs to $P(G,w)$. Moreover, any such $x=-\chi(C)/w(C)$, for a cycle $C$, is a vertex of $P(G,w)$ since there are $m$ linearly independent inequalities of $P(G,w)$ tight at $x$, namely: the conservation of flow equations $(F)$ at $|C|-1$ vertices of $C$, the equation $\sum_{e\in C}w_e y_e=-1$, and $m-|C|$ equations $y_e=0$, for $e\in E\setminus C$. 

To prove the opposite direction, let $y\in \RR^E$ be a vertex of $P(G,w)$. Let $Y=\{e\in E:~y_e>0\}$. The proof follows from the following 3 claims.

\begin{claim}\label{cl1}
The graph $(V,Y)$ is the disjoint union of strongly connected components.
\end{claim}
\proof
Consider an arbitrary strongly connected component $X$ in this graph, and let $X^-$ be the set of components reachable from $X$ (including $X$). Summing the conservation of flow equations corresponding to all the nodes in $X^-$ implies that all arcs going out of $X^-$ have a flow of zero. 
\qed


\begin{claim}\label{cl2} 
There exists no cycle $C\in\cC^0(G,w)$ such that $C\subseteq Y$.
\end{claim}  
\proof
If such a $C$ exists, we define two points $y'$ and $y''$ as follows. 
$$
y_e'=\left\{
\begin{array}{ll}
y_e+\epsilon,& \mbox{if $e\in C$}\\
y_e, &\mbox{otherwise,}
\end{array}
\right.
\ \ \ \ 
y_e''=\left\{
\begin{array}{ll}
y_e-\epsilon,& \mbox{if $e\in C$}\\
y_e, &\mbox{otherwise,}
\end{array}
\right.
$$
for some sufficiently small $\epsilon>0$. Then $y',y''$ clearly satisfy (F). Moreover, (N) is satisfied by $y'$ since 
$$\sum_{e\in E}w_e y'_e=\sum_{e\not\in C}w_ey_e+\sum_{e\in C}w_e(y_e+\epsilon)=\sum_{e\in E}w_ey_e+w(C)\epsilon=-1.$$
Similarly for $y''.$ Thus $y',y''\in P(G,w)$ and $y=(y'+y'')/2$ contradicting that $y$ is a vertex.
\qed


\begin{claim}\label{cl3} 
There exist no distinct cycles $C_1,C_2\in\cC^-(G,w)\cup\cC^+(G,w)$ such that $C_1\cup C_2\subseteq Y$.
\end{claim}  
\proof
If such $C_1$ and $C_2$ exist, we define two points $y'$ and $y''$ as follows. 
$$
y_e'=\left\{
\begin{array}{ll}
y_e+\epsilon_1,& \mbox{if $e\in C_1\setminus C_2$}\\
y_e+\epsilon_2,& \mbox{if $e\in C_2\setminus C_1$}\\
y_e+\epsilon_1+\epsilon_2,& \mbox{if $e\in C_1\cap C_2$}\\
y_e, &\mbox{otherwise,}
\end{array}
\right.
\ \ \ \ 
y_e'=\left\{
\begin{array}{ll}
y_e-\epsilon_1,& \mbox{if $e\in C_1\setminus C_2$}\\
y_e-\epsilon_2,& \mbox{if $e\in C_2\setminus C_1$}\\
y_e-\epsilon_1-\epsilon_2,& \mbox{if $e\in C_1\cap C_2$}\\
y_e, &\mbox{otherwise,}
\end{array}
\right.
$$
where $\epsilon_1=-\frac{w(C_2)}{w(C_1)}\epsilon_2$, for some sufficiently small $\epsilon_2>0$ (in particular, to insure non-negativity of $y',y''$, $\epsilon_2$ must be upper bounded by the minimum of $\min\{y_e: e\in C_2\setminus C_1\}$, $\frac{|w(C_1)|}{|w(C_2)|}\min\{y_e: e\in C_1\setminus C_2\}$, and $\frac{|w(C_1)|}{|w(C_1)-w(C_2)|}\min\{y_e: e\in C_1\cap C_2\}$). Then it is easy to verify that $y',y''$ satisfy (F). Moreover, (N) is satisfied with $y'$since 
\begin{eqnarray*}
\sum_{e\in E}w_e y'_e&=&\sum_{e\not\in C_1\cup C_2}w_ey_e+\sum_{e\in C_1\setminus C_2}w_e(y_e+\epsilon_1)+\sum_{e\in C_2\setminus C_1}w_e(y_e+\epsilon_2)\\
&&+\sum_{e\in C_1\cap C_2}w_e(y_e+\epsilon_1+\epsilon_2)=\sum_{e\in E}w_ey_e+w(C_1)\epsilon_1+w(C_2)\epsilon_2=-1.\end{eqnarray*}
Similarly for $y''.$ Thus $y',y''\in P(G,w)$ and $y=(y'+y'')/2$ contradicting that $y$ is a vertex of $P(G,w)$.
\qed


The above 3 claims imply that the graph $(V,Y)$ consists of a single cycle $C$ and a set of isolated vertices $V\setminus V(C)$. Thus $y_e=0$ for $e\not\in C$. By (F) we get that $y_e$ is the same for all $e\in C$, and by (N) we get that $y_e=-1/w(C)$ for all $e\in C$, and in particular that $C\in\cC^-(G,w)$. This completes the proof of Theorem \ref{t5}.

Since, in the graph $G$ arising from a $3$-CNF formula, every negative cycle has weight exactly $-1$, Theorem \ref{t5} implies that the vertices of $P(G,w)$ are exactly the characteristic vectors of the negative cycles of $G$. Recall that finding whether $G$ has any long negative cycle, \emph{i.e.} a negative cycle containing a marked edge (say $e$), is NP-complete. Further recall that for a $3$-CNF formula with $m$ clauses the constructed graph $G$ has $18m+1$ edges and $3m$ trivial short negative cycles. Consequently, the polyhedron $P(G,w)$ that is finally obtained has dimension $18m+1$ and $3m$ trivial vertices corresponding to the short negative cycles of $G$.

Now, if there are no long negative cycles then the vertex centroid of $P(G,w)$ has value $0$ in the coordinate corresponding to the edge $e$. For simplicity, we will refer to this coordinate axis as $x_e$. On the other hand, if there are $K\geq 1$ long negative cycles in $G$ then in the centroid $x_e=\frac{K}{K+3m}\geq \frac{1}{3m+1}$. This implies that having an $\epsilon$-approximation for the centroid of $P(G,w)$ for $\epsilon < \frac{1}{2(3m+1)}$ would reveal whether or not $P(G,w)$ has a non-trivial vertex and hence whether or not $G$ has a long negative cycle. Thus we have the following theorem:

\begin{theorem}\label{t6}
There is no polynomial algorithm that computes a $\frac{1}{d}$-approximation of the vertex centroid of an arbitrary $\mathcal{H}$-polyhedron $P\subset\RR^d$, unless $P=NP$.
\end{theorem}

An immediate consequence of Theorem \ref{t4} and Theorem \ref{t6} is that there is no polynomial algorithm that computes any ``sufficiently non-trivial'' approximation of the vertex centroid of an arbitrary $\mathcal{H}$-polyhedron unless $P=NP$. More formally,

\begin{corollary}
There is no polynomial algorithm that $d^{\frac{1}{2}-\delta}$-approximates the centroid of an arbitrary $d$-dimensional $\mathcal{H}$-polyhedron for any fixed constant $\delta>0$ unless $P=NP$.
\end{corollary}

\section{Open Problems}
Although we can show that for unbounded polyhedra almost any non-trivial approximation of the vertex centroid is hard, we can not make a similar statement for the bounded case (\emph{i.e. polytopes}. One interesting variant of Theorem \ref{t2} would be to consider a ball of radius $r$ instead of a halfspace. If containment of vertex centroid in a ball of radius $r$ can be decided in time polynomial in the number of inequalities defining the polytope, the dimension and $r$ then one can perform a sort of random walk inside the polytope and approximate the centroid in polynomial time. We leave out the details of this random walk since we do not have a method to check containment inside a ball.


%
%
%
%
%
%
{
\small
\bibliographystyle{abbrv}
\bibliography{centroid}}
\end{document}